\newcommand{\be}{\begin{equation}}
\newcommand{\ee}{\end{equation}}
\newcommand{\bear}{\begin{eqnarray}}
\newcommand{\eear}{\end{eqnarray}}
\newcommand{\ba}{\begin{array}}
\newcommand{\ea}{\end{array}}

\documentclass[preprint,showpacs,nofootinbib,preprintnumbers,amsmath,amssymb,superscriptaddress]{revtex4}
\usepackage{epsfig,graphicx,psfrag}

\begin{document}

\preprint{} \ \ \

\title{\Large \bf $A_{4}$ See-Saw Models and Form Dominance}

\author{Mu-Chun Chen}
\affiliation{Department of Physics \& Astronomy\\ University of California, Irvine, CA 92697-4575, U.S.A.}

\author{Stephen F. King}
\affiliation{School of Physics \& Astronomy\\ University of Southampton, Southampton, SO17 1BJ, U.K.}

\date{
1 March 2009 \rule{0em}{1.8em}}

\begin{abstract}\rule{0em}{1.8em}
We introduce the idea of Form Dominance in the (type I) see-saw mechanism,
according to which a particular right-handed neutrino
mass eigenstate is associated with a particular physical neutrino mass eigenstate,
leading to a form diagonalizable effective neutrino mass matrix.
Form Dominance, which allows an arbitrary neutrino mass spectrum,
may be regarded as a generalization of Constrained Sequential
Dominance which only allows strongly hierarchical neutrino masses.
We consider alternative implementations of the see-saw mechanism in minimal $A_{4}$ see-saw models
and show that such models satisfy Form Dominance,
leading to neutrino mass sum rules which predict closely spaced neutrino masses
with a normal or inverted neutrino mass ordering. To avoid the partial cancellations inherent in such models
we propose Natural Form Dominance, in which a different flavon is associated with each physical neutrino mass eigenstate.
\end{abstract}


\maketitle

\section{Introduction}
The most remarkable discovery in particle physics over the past
decade has been the discovery of neutrino mass and mixing
involving two large mixing angles commonly known as the
atmospheric angle $\theta_{23}$ and the solar angle $\theta_{12}$.
The latest data from neutrino oscillation experiments is
consistent with the so called tri-bimaximal mixing (TBM) mixing
pattern \cite{tribi},
\begin{equation}
\label{TBM}
U_{TBM}= \left(\begin{array}{ccc} -\frac{2}{\sqrt{6}}& \frac{1}{\sqrt{3}}&0\\
\frac{1}{\sqrt{6}}&\frac{1}{\sqrt{3}}&\frac{1}{\sqrt{2}}\\
\frac{1}{\sqrt{6}}&\frac{1}{\sqrt{3}}&-\frac{1}{\sqrt{2}}
\end{array} \right).
\end{equation}
The question of how to achieve TBM has been the subject of intense
theoretical speculation and
there have been many attempts to
derive TBM from models based on an underlying family symmetry
spontaneously broken by new Higgs fields called ``flavons''
\cite{Frampton:2004ud,Ma:2007wu,Altarelli:2006kg,Feruglio:2007uu,
deMedeirosVarzielas:2005ax,King:2006np,King:2006me,km,Chen:2007af,
Mohapatra:2006pu,Chan:2007ng}.
Since the forthcoming
neutrino experiments will be sensitive to small deviations from
TBM, it is important to study the theoretical uncertainty in such
TBM predictions, and this has also been addressed \cite{Antusch:2008yc}.

Although in the above theoretical models \cite{Frampton:2004ud,Ma:2007wu,Altarelli:2006kg,Feruglio:2007uu,
deMedeirosVarzielas:2005ax,King:2006np,King:2006me,km,Chen:2007af,
Mohapatra:2006pu,Chan:2007ng} the neutrino and charged lepton mass matrices
are always constructed in some particular basis, the physical results must always be basis invariant.
For example, models of TBM based on the discrete family symmetry group $A_4$
were originally constructed in a basis
in which both the neutrino and charged lepton mass
matrices are both non-diagonal \cite{Ma:2007wu},
but were subsequently reformulated in the more convenient
flavour basis in which the charged lepton masses were diagonal
\cite{Altarelli:2006kg}. Similarly, when the see-saw mechanism is considered,
the Dirac neutrino and right-handed neutrino Majorana mass matrices are also constructed in a
particular basis but again the results must be basis invariant.
Thus, although see-saw models of TBM based on the discrete family symmetry group $A_4$
have so far been constructed in a basis in which the right-handed neutrino mass matrix is not diagonal~\cite{Altarelli:2006kg},
in this paper we shall find it convenient to
consider such models in the diagonal right-handed neutrino mass basis.

In this paper we introduce the idea of Form Dominance (FD) in the (type I) see-saw mechanism
as a generic and natural mechanism which
leads to a form diagonalizable effective neutrino mass matrix
in which the mixing matrix is independent of the parameters which control the
physical neutrino masses. It is well known that models
which reproduce a form diagonalizable effective neutrino mass matrix can
provide a natural explanation of TBM without any fine-tuning of parameters \cite{Low:2003dz}.
Here we shall show how to achieve such a form diagonalizable effective neutrino mass matrix starting
from the type I see-saw mechanism using the FD mechanism. The basic idea of
FD is that a particular right-handed neutrino mass eigenstate
is associated with a particular physical neutrino mass eigenstate,
similar to the case of Constrained Sequential Dominance (CSD) \cite{King:2006np,King:1998jw}.
However, whereas CSD only applies to the case of a strong neutrino mass hierarchy,
FD is more general and allows three physical neutrino masses with arbitrary masses and ordering.
As an example of FD we shall consider minimal $A_{4}$ see-saw models, in which the neutrino sector
involves only one triplet flavon plus one singlet flavon, including both the usual see-saw model
proposed in \cite{Altarelli:2006kg} and a new alternative one.
Working in the diagonal right-handed neutrino mass basis, we shall show that
both these see-saw models satisfy FD
leading to neutrino mass sum rules which predict closely spaced neutrino masses
with a normal or inverted neutrino mass ordering~\cite{Chen:2007af}.
The results motivate the idea of Natural Form Dominance (NFD), in which a different flavon is
associated with each physical neutrino mass eigenstate, with CSD as a special case
of NFD.

The remainder of the paper is organized as follows. In section \ref{II} we give the form
of the TBM effective neutrino mass matrix in the flavour basis where it may be expressed
in terms of neutrino masses and columns of the MNS matrix. In section \ref{III} we introduce the
idea of FD in a particular basis, then in a basis invariant way.
In section \ref{IV} we examine the minimal $A_4$ see-saw models defined above in the diagonal
right-handed neutrino mass basis (not usually considered) and
show that they satisfy FD, with the neutrino masses obeying various sum rules corresponding to
closely spaced neutrino masses
with a normal or inverted neutrino mass ordering.
Section \ref{V} is reserved for a discussion of our results,
including the motivation for NFD, and the conclusion.

\section{TBM in the Flavour Basis \label{II}}
In the flavour basis, in which the charged lepton mass matrix is
diagonal and the TBM arises from the neutrino sector, the effective neutrino
mass matrix corresponding to TBM, denoted by $({M^{\nu}_{eff}})^{TBM}$,
may be diagonalized as,
\begin{equation}
{M^{\nu}_{eff}}^{\mbox{\scriptsize diag}} = U_{\mbox{\scriptsize TBM}}^{T}
({M^{\nu}_{eff}})^{TBM} U_{\mbox{\scriptsize TBM}}=(m_{1}, \; m_{2}, \; m_{3}) \; .
\end{equation}
Given $U_{\mbox{\scriptsize TBM}}$, this
enables $({M^{\nu}_{eff}})^{TBM}$ to be determined in terms of neutrino masses,
\begin{equation}\label{eq:csd-tbm0}
({M^{\nu}_{eff}})^{TBM}= m_{1} \Phi_{1}\Phi_{1}^{T} + m_{2} \Phi_{2}\Phi_{2}^{T} + m_{3} \Phi_{3}\Phi_{3}^{T} \; ,
\end{equation}
where the three matrices are
\begin{equation}\label{eq:csd-tbmdec0}
\Phi_{1} \Phi_{1}^{T} = \frac{1}{6}
\left(\begin{array}{ccc}
4 & -2 & -2 \\
-2 & 1 & 1 \\
-2 & 1 & 1
\end{array}\right), \;
\Phi_{2} \Phi_{2}^{T} = \frac{1}{3}
\left(\begin{array}{ccc}
1 & 1 & 1 \\
1 & 1 & 1 \\
1 & 1 & 1
\end{array}\right), \;
\Phi_{3}\Phi_{3}^{T} = \frac{1}{2}\left(\begin{array}{ccc}
0 & 0 & 0 \\
0 & 1 & -1 \\
0 & -1 & 1
\end{array}\right) \; ,
\end{equation}
corresponding to the orthonormal column vectors
\begin{equation}
\label{Phi0}
{\Phi}_{1}=\frac{1}{\sqrt{6}}
\left(
\begin{array}{c}
-2 \\
1 \\
1
\end{array}
\right), \ \
{\Phi}_{2}=\frac{1}{\sqrt{3}}
\left(
\begin{array}{c}
1 \\
1 \\
1
\end{array}
\right), \ \
{\Phi}_{3}=\frac{1}{\sqrt{2}}
\left(
\begin{array}{c}
0 \\
1 \\
-1
\end{array}
\right).
\end{equation}
Note that $\Phi_{1,2,3}$ are just the three columns of $U_{\mbox{\scriptsize TBM}}$,
namely,
\begin{equation}
\Phi_{1i}={U_{\mbox{\scriptsize TBM}}}_{i1},\ \
\Phi_{2i}={U_{\mbox{\scriptsize TBM}}}_{i2},\ \
\Phi_{3i}={U_{\mbox{\scriptsize TBM}}}_{i3}.
\end{equation}

From above we may write $({M^{\nu}_{eff}})^{TBM}$ as the symmetric matrix,
\begin{equation}\label{eq:csd-tbm2}
({M^{\nu}_{eff}})^{TBM}=
\left(\begin{array}{ccc}
a & b & c \\
. & d & e \\
. & . & f
\end{array}\right),
\end{equation}
where,
\begin{eqnarray}\label{abc}
a &=& \frac{2}{3}m_1+\frac{1}{3}m_2,\nonumber \\
b &=& c=-\frac{1}{3}m_1+\frac{1}{3}m_2,\nonumber \\
d &=& f=\frac{1}{6}m_1+\frac{1}{3}m_2 +\frac{1}{2}m_3,\nonumber \\
e &=& a+b-d.
\end{eqnarray}
In particular $b=c$ and $d=f$ and $e = a+b-d$
are the characteristic signatures of the TBM neutrino mass matrix in the flavour basis.

\section{Form Dominance \label{III}}
The key requirement of a form diagonalizable effective neutrino mass matrix
is the presence of no more than three free parameters in the matrix, which are subsequently  related to the
physical neutrino mass eigenvalues. For example a form diagonalizable effective neutrino mass matrix
in the notation of Eq.\ref{eq:csd-tbm2} involving only three free parameters $a,b,d$ and taking the form
\begin{equation}\label{eq:csd-tbm3}
({M^{\nu}_{eff}})^{TBM} =
\left(\begin{array}{ccc}
\ \ \ a \ \ \ & \ \ \ b \ \ \  & b \\
. & d & (a+b-d) \\
. & . & d
\end{array}\right),
\end{equation}
will result in TBM independently of the parameters $a,b,d$
and hence independently of the physical neutrino masses which are related to
the parameters $a,b,d$ by Eq.\ref{abc}. On the other hand if there are
more than three free parameters then one or more of the conditions
in Eq.\ref{abc} $b=c$ and $d=f$ and $e = a+b-d$
would have to be achieved by tuning and such a matrix would then not be form diagonalizable
since the mixing matrix would depend on the parameter choice (and hence depend on the
physical neutrino masses). It is clear that the notion of a form diagonalizable effective neutrino
mass matrix is related to its dependence on only three (or less) free parameters. If this matrix arises from the
type I see-saw mechanism, it is not {\it a priori} obvious how the underlying theory, involving the
Dirac neutrino mass matrix and heavy Majorana neutrino mass matrix could naturally lead to such a form diagonalizable
effective neutrino mass matrix. In general,
the see-saw mechanism involves many parameters which could enter
into the effective neutrino mass matrix arising from
the non-symmetric Dirac mass matrix as well as the (typically) three right-handed neutrino masses.
To achieve a form diagonalizable effective neutrino mass matrix it is clearly necessary to constrain the form
of the Dirac neutrino mass matrix, and also associate the right-handed neutrino masses with the Dirac mass matrix,
in such a way that only three (or fewer) independent combinations of parameters enter the effective neutrino mass matrix.

Constrained Sequential Dominance (CSD)
\cite{King:2006np,King:1998jw} provides an example of how this may
be achieved for the case of strongly hierarchical neutrino masses.
According to CSD, in the diagonal right-handed neutrino mass
basis, each column of the Dirac mass matrix is associated with a
particular right-handed neutrino mass, and CSD then imposes the
constraint that these columns are proportional to those in
Eq.\ref{Phi0}, leading to only three independent parameters
entering the effective neutrino mass matrix, which is consequently
form diagonalizable. However CSD assumes a strong physical
neutrino mass hierarchy, $|m_1|\ll |m_2|< |m_3|$ so that
effectively the subdominant column associated with $m_1$ may be
neglected, and then only two free parameters associated with two
right-handed neutrinos responsible for $m_2$ and $m_3$ remain
\cite{King:2006np}. Here we shall discuss a generalization of CSD
applicable to the case of three physical neutrino masses $m_1,
m_2, m_3$ with arbitrary mass values and mass orderings (including
the cases of an inverted hierarchy and quasi-degenerate neutrinos
as well as hierarchical neutrinos). In other words we shall propose a more
general framework which has all the nice properties of CSD, but
which allows a non-hierarchical neutrino mass spectrum.

We now introduce the notion of
Form Dominance (FD) in the type I see-saw mechanism as an elegant and generic mechanism for achieving
a form diagonalizable effective neutrino mass matrix from the type I see-saw mechanism.
FD may be defined in the diagonal right-handed neutrino mass basis and diagonal
charged lepton mass basis as follows.
To set the notation, recall that,
in the type I see-saw mechanism, the starting point is a heavy right-handed Majorana neutrino mass matrix
$M_{RR}$ and a Dirac neutrino mass matrix (in the left-right convention) $M_{D}$, with the light effective left-handed Majorana
neutrino mass matrix $M_{eff}^{\nu}$ given by the type I see-saw formula \cite{Minkowski:1977sc},
\begin{equation}\label{eq:meff}
M_{eff}^{\nu} = M_{D} M_{RR}^{-1} M_{D}^{T}.
\end{equation}
In a basis in which $M_{RR}$ is diagonal, we may write,
\begin{equation}
M_{RR} = \mbox{diag}(M_A, M_B, M_C)
\end{equation}
and $M_{D}$ may be written in terms of three general column vectors $A,B,C$,
\begin{equation}
M_{D} = (A,B,C).
\end{equation}
The see-saw formula then gives,
\begin{equation}
\label{eq:seesawmeff}
M_{eff}^{\nu} =
 \frac{AA^{T}}{M_A} + \frac{BB^{T}}{M_{B}} +\frac{CC^{T}}{M_{C}}.
\end{equation}
Using this notation, FD may now be defined as follows.
FD is the requirement that each column of the Dirac mass
matrix (in the particular basis defined above) is proportional to a different column of the MNS matrix $U$,
\begin{equation}
\label{FD}
A_i =a U_{i1},\ \
B_i =b U_{i2},\ \
C_i =c U_{i3}.
\end{equation}
It is then clear that, although there are six parameters in the see-saw theory, only three
independent combinations will enter into the effective neutrino mass matrix, after the type I
see-saw mechanism. To be precise,
the three constants of proportionality $a,b,c$ in Eq.\ref{FD} combine
with the three right-handed neutrino masses $M_{A,B,C}$ to yield three independent combinations of parameters
appearing in the effective neutrino mass matrix $M_{eff}^{\nu}$ given by the see-saw mechanism in
Eq.\ref{eq:seesawmeff}. Moreover, the resulting $M_{eff}^{\nu}$ is form diagonalizable,
diagonalized by the MNS matrix $U$, with
the physical neutrino masses $m_i$ given by $a^2/M_A$, $b^2/M_B$, $c^2/M_C$.
In such a case, each right-handed neutrino mass eigenstate is clearly associated with a particular
physical neutrino mass eigenstate of mass $m_i$. We emphasize that FD applies to any general
MNS mixing matrix $U$, not just TBM.

The notion of FD may now simply be applied to the special case of TBM,
in the particular basis defined above, namely the diagonal
right-handed neutrino mass basis and diagonal charged lepton mass
basis. Applying the FD conditions in Eq.\ref{FD} to the case of
$U=U_{\mbox{\scriptsize TBM}}$, by comparing
Eq.\ref{eq:seesawmeff} to
Eqs.\ref{eq:csd-tbm0},\ref{eq:csd-tbmdec0}, \ref{Phi0} it is clear
that an effective neutrino mass matrix of the TBM type
$({M^{\nu}_{eff}})^{TBM}$ may be achieved if
\begin{equation}
\label{FD2}
A = a\Phi_1 = \frac{a}{\sqrt{6}}
\left(
\begin{array}{c}
-2 \\
1 \\
1
\end{array}
\right),\ \
B = b \Phi_2 = \frac{b}{\sqrt{3}}
\left(
\begin{array}{c}
1 \\
1 \\
1
\end{array}
\right),\ \
C = c \Phi_3 = \frac{c}{\sqrt{2}}
\left(
\begin{array}{c}
0 \\
1 \\
-1
\end{array}
\right).
\end{equation}
Moreover, the constraints in Eq.\ref{FD2} lead to a form diagonalizable $({M^{\nu}_{eff}})^{TBM}$
diagonalized by $U_{\mbox{\scriptsize TBM}}$ (in this basis) with physical neutrino
mass eigenvalues given by $m_1=a^2/M_A$, $m_2=b^2/M_B$, $m_3=c^2/M_C$, as indicated previously.

It is interesting to compare FD for TBM defined above to
Constrained Sequential Dominance (CSD) defined in
\cite{King:2006np,King:1998jw}. In CSD a strong hierarchy $|m_1|\ll |m_2| < |m_3|$ is assumed
which enables $m_1$ to be effectively ignored (typically this is achieved by
taking $M_A$ to be very heavy leading to a very light $m_1$) then CSD is defined by
only assuming the second and third conditions in Eq.\ref{FD2} \cite{King:2006np}.
Thus CSD is seen to be just a special case of FD corresponding to a strong neutrino mass
hierarchy. FD on the other hand is more general
and allows any choice of neutrino masses including
a mild hierarchy, an inverted hierarchy or a quasi-degenerate mass pattern.

Finally note that FD can also be defined in a basis invariant way as follows. In a
general basis, one can always write, without loss of generality
\begin{equation}
\left(\begin{array}{ccc} A_iM_A^{-1/2} & B_iM_B^{-1/2} &
C_iM_C^{-1/2}
\end{array}\right)
= \left(\begin{array}{ccc} U_{i1}{m}_1^{1/2}& U_{i2}{m}_2^{1/2} &
U_{i3}{m}_3^{1/2}
\end{array}\right)R^T
\label{R}
\end{equation}
where $R$ is the general orthogonal matrix introduced by Casas and
Ibarra \cite{Casas:2001sr}. Then FD corresponds to the case of $R$
equal to the unit matrix (up to permutations which just
corresponds to a relabelling of $A,B,C$). This was in fact
observed in \cite{King:2006hn} \footnote{See Eq.92 of
\cite{King:2006hn} and the subsequent discussion. Note also that
to be fully general the column vectors introduced in Eq.\ref{Phi0}
should have been post multiplied by $R^T$.} but only the limit
$m_1=0$ was considered corresponding to CSD, where it was noted
that the $R$ matrix formalism provides a basis invariant
formulation of CSD (since $R$ is basis invariant). Here we allow
for general $m_{i}$ with FD defined by $R$ equal to the unit
matrix providing a basis invariant definition of FD. However in
practice we shall work in the diagonal right-handed neutrino mass
basis and diagonal charged lepton mass basis discussed previously.

\section{Minimal $A_4$ Models and Form Dominance \label{IV}}

In this section we discuss Form Dominance (FD) in the framework
of minimal $A_4$ models, where minimal means that the neutrino sector only involves
one triplet flavon plus one singlet flavon. We shall work in the diagonal charged lepton mass
basis (referred to as the flavour basis). To be as general as possible, we discuss these models independently
of a particular mechanism for vacuum alignment or of the symmetries required to enforce the
operator structure of the models.

\subsection{Minimal $A_4$ Models}
The group $A_{4}$ is a group that describes even permutations of four objects. It has two generators, $S$ and $T$, and four
inequivalent irreducible representations, $1$, $1^{\prime}$, $1^{\prime\prime}$ and $3$.
In the diagonal basis for $T$, the two generators are given in the triplet representation as,
\begin{equation}
S = \frac{1}{3} \left(\begin{array}{ccc}
-1& 2  & 2  \\
2  & -1  & 2 \\
2 & 2 & -1
\end{array}\right), \quad
T = \left( \begin{array}{ccc}
1 & 0 & 0 \\
0 & \omega^{2} & 0 \\
0 & 0 & \omega
\end{array}\right) \; ,
\end{equation}
where $\omega = e^{2\pi i/3}$. The product rules are
$3 \times 3 = 1 + 1^{\prime} + 1^{\prime\prime} + 3_{S} + 3_{A}$, and $1^{a} \times 1^{b} = 1^{(a+b) \; \mbox{mod} \; 3}$, where $a, \; b=0, \; 1, \; 2$ for representation $1$, $1^{\prime}$ and $1^{\prime\prime}$, respectively. The Clebsch-Gordon coefficients of the above product rules can be found in Ref.~\cite{Altarelli:2006kg}.

Without the right-handed neutrinos, the small neutrino masses can
be generated by the dimension-5 operator which breaks both the
total and individual lepton numbers,
\begin{equation} y_{ij}
\frac{\overline{\ell_{i}^{c}} \ell_{j} HH}{\Lambda_{L}} \; ,
\end{equation}
where
$\ell_{i}$ ($i=1,2,3$) are the lepton doublets, $H$ is the SM
Higgs, $y_{ij}$ are the Yukawa couplings and $\Lambda_{L}$ is the cutoff scale for the lepton number violation operator. The tri-bimaximal
mixing pattern arises if the three lepton doublets transform as a
triplet of $A_{4}$, and the three right-handed charged leptons are
assigned to be singlets under $A_{4}$,
\begin{equation}
L =
\left(\begin{array}{c} \ell_{1} \\ \ell_{2} \\ \ell_{3}
\end{array}\right) \sim 3  \; ,  \quad e_{R} \sim 1 \; ,
\quad \mu_{R} \sim 1^{\prime\prime} \; , \quad  \tau_{R} \sim 1^{\prime} \;.
\end{equation}
The Lagrangian that gives rise to neutrino masses
is
\begin{equation} \mathcal{L}_{LL} = \frac{\overline{L^{c}}LHH}{\Lambda_{L}}
\left(\frac{\left< \phi_{S} \right>}{\Lambda} + \frac{\left< u
\right>}{\Lambda}\right) \; ,
\end{equation}
where $\Lambda$ is the cutoff scale of the $A_{4}$ symmetry. The triplet
flavon field, $\phi_{S} \sim 3$, and the singlet flavon field, $u \sim
1$, acquire the following vacuum expectation values (VEVs),
\begin{equation}
\frac{\left< \phi_{S} \right>}{\Lambda} =
\left(\begin{array}{c} 1 \\ 1 \\ 1 \end{array}\right) \alpha_{s}  \; ,
\quad \frac{\left< u \right>}{\Lambda} = \alpha_{0}  \; .
\end{equation}
The VEV $\left<\phi_{S}\right>$ breaks the $A_{4}$ symmetry down to $G_{S}$,
which is the subgroup of $A_{4}$ generated by the group element $S$.
Upon the electroweak symmetry breaking, the following effective
neutrino mass matrix is generated,
\begin{equation}\label{mr}
{M^{\nu}_{eff}} = \left( \begin{array}{ccc} 2\alpha_{s} + \alpha_{0} & -\alpha_{s} & -\alpha_{s} \\ -\alpha_{s} & 2\alpha_{s} &
-\alpha_{s} + \alpha_{0} \\ -\alpha_{s} & -\alpha_{s} + \alpha_{0} & 2\alpha_{s} \end{array}\right)
\frac{v^{2}}{\Lambda_{L}}\; ,
\end{equation}
where $v$ is the SM
Higgs VEV. This mass matrix is form-diagonalizable, {\it i.e.} it
is always diagonalized, independent of the values for the
parameters $\alpha_{s}$ and $\alpha_{0}$,  by the tri-bimaximal mixing matrix,
\begin{equation}
{M^{\nu}_{eff}}^{\mbox{\scriptsize diag}} =
U_{\mbox{\scriptsize TBM}}^{T} M^{\nu}_{eff} U_{\mbox{\scriptsize TBM}}
= \mbox{diag}(3\alpha_{s}+\alpha_{0}, \; \alpha_{0}, \; 3\alpha_{s}-\alpha_{0}) \cdot \frac{v^{2}}{\Lambda_{L}}
\equiv (m_{1}, \; m_{2}, \; m_{3}) \; .
\end{equation}
Because the three mass eigenvalues $m_{1,2,3}$ are determined by two
parameters, $a$ and $b$, there is a sum rule among the three light
masses~\cite{Chen:2007af},
\begin{equation}
m_{1} - m_{3} = 2 m_{2} \; .
\end{equation}
Given that the solar mass squared
difference is positive, this sum rule leads to a prediction for
the normal mass hierarchy in the atmospheric neutrino sector. The
charged lepton masses are generated due to the following
Lagrangian,
\begin{equation}\label{eq:chglep}
\mathcal{L}_{\mbox{\scriptsize lep}} = \frac{1}{\Lambda} \left(y_{e} (\overline{\ell}
\phi^{\prime})_{1} e_{R} H + y_{\mu} (\overline{\ell}
\phi^{\prime})_{1^{\prime}} \mu_{R} H + y_{\tau} (\overline{\ell}
\phi^{\prime})_{1^{\prime\prime}} \tau_{R} H \right) \; .
\end{equation}
Here the
triplet flavon field, $\phi_{T}$, acquires a VEV along the
following direction,
\begin{equation} \left< \phi_{T} \right>
= \left( \begin{array}{c} v_{T} \\ 0 \\ 0 \end{array} \right)
\; ,
\end{equation}
breaking the $A_{4}$ symmetry down to $G_{T}$, which is the subgroup generated by $T$.
This leads to a diagonal charged lepton mass
matrix, {\it i.e.} $V_{e,L} = I$, and thus the PMNS matrix is
exactly of the tri-bimaximal form, $U_{PMNS} = V_{e,L}V_{\nu}^{\dagger} = U_{TBM}$.

\subsection{The See-saw Mechanism and Form Dominance in Minimal $A_4$ Models}
The minimal $A_4$ see-saw realization of the tri-bimaximal mixing pattern has
been discussed before~\cite{Altarelli:2006kg}, however here
we examine such models in the diagonal right-handed neutrino mass
basis, which has not been discussed before.
We discuss two alternative see-saw realizations of
the tri-bimaximal mixing pattern in the basis where the charged
lepton mass matrix is diagonal $V_{e, L} = I$,
the usual one proposed in~\cite{Altarelli:2006kg},
and an alternative example which we propose. We shall show that in
both examples the FD mechanism is present.

\subsubsection{The usual see-saw realization}
Since the charged lepton mass matrix is diagonal in our
realization, it is generated by the same Lagrangian as given above
in Eq.~\ref{eq:chglep}.  The three right-handed neutrinos
transform as a triplet of $A_{4}$,
\begin{equation}
N = \left(\begin{array}{c}
N_{1} \\ N_{2} \\ N_{3}
\end{array}\right) \sim 3 \; ,
\end{equation}
and the right-handed neutrino Majorana mass matrix is generated by,
\begin{equation}\label{eq:mr-1}
M_{RR} =  \overline{N^{c}} N ( \left< \phi_{S} \right> + \left< u \right>)
= \left( \begin{array}{ccc}
2\alpha_{s} + \alpha_{0} & -\alpha_{s} & -\alpha_{s} \\
-\alpha_{s} & 2\alpha_{s} & -\alpha_{s} + \alpha_{0} \\
-\alpha_{s} & -\alpha_{s} + \alpha_{0} & 2\alpha_{s}
\end{array}\right) \Lambda \; .
\end{equation}
The Dirac neutrino mass matrix is generated by the following interaction,
\begin{equation}\label{eq:md-1}
M_{D} = y H \overline{L} N = \left(\begin{array}{ccc}
1 & 0 & 0
\\
0 & 0 & 1
\\
0 & 1 & 0
\end{array}\right) yv \; .
\end{equation}
After the see-saw mechanism takes place, the resulting effective neutrino mass matrix is
\begin{equation}
M_{\nu}^{eff} =  M_{D} M_{RR}^{-1} M_{D}^{T} = U_{TBM}^{T} \mbox{diag}(m_{1}, m_{2}, m_{3}) U_{TBM}.
\end{equation}
This effective neutrino mass matrix is diagonalized by $U_{TBM}$ with the mass eigenvalues being
\begin{equation}
\label{m}
\mbox{diag}(m_{1}, m_{2}, m_{3}) = \left(
\frac{1}{3\alpha_{s}+\alpha_{0}}, \; \frac{1}{\alpha_{0}}, \; \frac{1}{3\alpha_{s}-\alpha_{0}}\right) \frac{y^{2} v^{2}}{\Lambda} \; .
\end{equation}
The sum rule among the three light neutrino masses is given in this see-saw realization by,
\begin{equation}
\frac{1}{m_{1}} - \frac{1}{m_{3}} = \frac{2}{m_{2}} \; ,
\end{equation}
which can lead to both normal and inverted hierarchical mass orderings.

In the see-saw realization of the tri-bimaximal mixing described above, the right-handed neutrino Majorana mass matrix $M_{RR}$ is diagonalized by
\begin{equation}
\label{mrr}
M_{RR}^{diag}  = U_{TBM}^{T} M_{RR} U_{TBM} =
\mbox{diag}(3\alpha_{s}+\alpha_{0}, \; \alpha_{0}, \; 3\alpha_{s}-\alpha_{0}) \Lambda \equiv \mbox{diag}(M_{A}, M_{B}, M_{C}).
\end{equation}
Rotating to the diagonal basis for the right-handed neutrino mass matrix $M_{RR}$, the Dirac mass matrix is given by,
\begin{equation}
\label{eq:Md1}
M_{D}^{\prime} = M_{D} U_{TBM}
= y v
\left(\begin{array}{ccc} -\frac{2}{\sqrt{6}}& \frac{1}{\sqrt{3}}&0\\
\frac{1}{\sqrt{6}}&\frac{1}{\sqrt{3}}&- \frac{1}{\sqrt{2}}\\
\frac{1}{\sqrt{6}}&\frac{1}{\sqrt{3}}& \frac{1}{\sqrt{2}}
\end{array}\right)
\equiv ( A, \; B, \; C) \; ,
\end{equation}
where $A$, $B$ and $C$ are three column vectors of $M_{D}^{\prime}$.
Comparing Eq.~\ref{eq:Md1} and Eq.~\ref{FD2} we see that Form Dominance is satisfied
with the proportionality constants being, $a=b= - c=y v$, so that in this basis the Dirac
mass matrix is in fact exactly proportional to the TBM mixing matrix.
\footnote{This is a special case. In general FD does not require that the Dirac mass matrix
be proportional to the MNS matrix, only that the respective columns be proportional.}
Form Dominance is thus at work in this model.

The physical light neutrino masses
are given by Eq.\ref{m}, with the mass splittings being controlled by the right-handed
neutrino masses in Eq.\ref{mrr}, $M_A = 3 \alpha_s + \alpha_0$
$M_B = \alpha_0$ and $M_C = 3 \alpha_s - \alpha_0$. Since these masses are controlled
by linear combinations of two VEVs $\alpha_0$ and $\alpha_s$, some partial cancellations are required to obtain
an acceptable neutrino mass pattern and
it is impossible to obtain a strong neutrino mass hierarchy in this model.

\subsubsection{An alternative see-saw realization}
Alternatively, the see-saw mechanism can be implemented in the following way\footnote{For implementation in the diagonal basis for $S$, see \cite{Hirsch:2008rp}.}.
Instead of the interactions given in Eq.~\ref{eq:mr-1} and \ref{eq:md-1}, consider that the Dirac mass term is generated by
\begin{equation}
M_{D} =  H \overline{L} N \left( \frac{\left< \phi_{S} \right>}{\Lambda} + \frac{\left< u \right>}{\Lambda} \right)
= \left( \begin{array}{ccc}
2\alpha_{s} + \alpha_{0} & -\alpha_{s} & -\alpha_{s} \\
-\alpha_{s} & 2\alpha_{s} & -\alpha_{s} + \alpha_{0} \\
-\alpha_{s} & -\alpha_{s} + \alpha_{0} & 2\alpha_{s}
\end{array}\right) v \; ,
\end{equation}
and the Majorana mass matrix is generated by,
\begin{equation}\label{md}
M_{RR} =  M_{R} \overline{N^{c}} N  = \left(\begin{array}{ccc}
1 & 0 & 0
\\
0 & 0 & 1
\\
0 & 1 & 0
\end{array}\right) M_{R} \; .
\end{equation}
Similar to the previous consideration, the charged lepton mass matrix is generated by the Lagrangian given in Eq.~\ref{eq:chglep}, and thus it is diagonal.
In this case, one can easily check that the neutrino mixing matrix is also of the tri-bimaximal form and the three effective mass eigenvalues are,
\begin{equation}
(m_{1}, \; m_{2}, \; m_{3}) = \biggl(
(3\alpha_{s}+\alpha_{0})^{2}, \; \alpha_{0}^{2}, \;
(3\alpha_{s}-\alpha_{0})^{2} \biggr) \frac{v^{2}}{M_{R}} \;  ,
\end{equation}
leading to a mass sum rule,
\begin{eqnarray}
\bigl| |\sqrt{m_{1}}| - |\sqrt{m_{3}}| \bigr| = 2 |\sqrt{m_{2}}| \; , \quad \mbox{for} \quad (3\alpha_{s} + \alpha_{0})(3\alpha_{s}-\alpha_{0})>0 \\
\bigl| |\sqrt{m_{1}} |+ |\sqrt{m_{3}}| \bigr| = 2 |\sqrt{m_{2}}| \; , \quad \mbox{for} \quad (3\alpha_{s} +\alpha_{0})(3\alpha_{s}-\alpha_{0})<0 \; .
\end{eqnarray}
In this see-saw realization, the RH neutrino mass matrix can be diagonalized by the TBM matrix,
\begin{equation}
U_{TBM}^{T} M_{RR} U_{TBM} = \mbox{diag} (1, 1, -1) M_{R} \; .
\end{equation}
In this basis, the Dirac neutrino mass matrix becomes,
 \begin{equation}
M_{D}^{\prime} =
\left(\begin{array}{ccc} -\frac{2}{\sqrt{6}}& \frac{1}{\sqrt{3}}&0\\
\frac{1}{\sqrt{6}}&\frac{1}{\sqrt{3}}&\frac{1}{\sqrt{2}}\\
\frac{1}{\sqrt{6}}&\frac{1}{\sqrt{3}}&-\frac{1}{\sqrt{2}}
\end{array}\right)
\cdot
\mbox{diag}(3\alpha_{s}+\alpha_{0}, \; \alpha_{0}, \; 3\alpha_{s}-\alpha_{0})
v \equiv (A,B,C) \; ,
\end{equation}
Similar to the previous case, there exists a correspondence between the vectors $A, \; B, \; C$  and $\Phi_{1,2,3}$ with proportionality constants being
\begin{equation}
a = 3\alpha_{s}+\alpha_{0}, \; b = \alpha_{0}, \; c=3\alpha_{s}-\alpha_{0} \; ,
\end{equation}
and thus the FD mechanism is at work. In this case the right-handed neutrino masses are degenerate,
and the physical light neutrino masses are proportional to $a^2,b^2,c^2$.
Since $a,b,c$ are controlled
by linear combinations of two VEVs $\alpha_0$ and $\alpha_s$, as before,
some partial cancellations are required to obtain
an acceptable neutrino mass pattern
and it is again impossible to obtain a strong neutrino mass hierarchy in this model.

\section{Discussion and Conclusion\label{V}}
The current experimental best fit values for the neutrino mixing
angles indicate that the neutrino mixing matrix resembles
the TBM form. This suggests an underlying (possibly discrete) family symmetry,
such as, for example, $A_4$. In the most attractive such models,
the resulting effective neutrino mass matrix is form diagonalizable,
that is to say it involves three (or less) parameters and,
in the diagonal charged lepton mass basis, is diagonalized by the TBM matrix,
independently of the choice of the parameters.

In this paper we have considered in general terms how such a form diagonalizable effective
neutrino mass matrix could result from the type I see-saw mechanism.
Clearly a necessary condition is that
the type I see-saw mechanism must lead to only three (or less) independent parameters
appearing in the effective mass matrix. We have proposed FD as an elegant way
to achieve a form diagonalizable effective neutrino mass matrix,
starting from the type I see-saw mechanism.
According to FD a particular right-handed neutrino
mass eigenstate is associated with a particular physical neutrino mass eigenstate in such a way that, in the
diagonal right-handed neutrino mass basis, the columns of the Dirac mass matrix are proportional to columns
of the MNS matrix, as in Eq.\ref{FD} in general, or Eq.\ref{FD2} for the TBM case.
FD may be regarded as a generalization of CSD,
but whereas CSD assumes a strong neutrino mass hierarchy, FD allows
three physical neutrino masses with arbitrary masses and ordering.

As an example of these ideas we have considered minimal $A_{4}$ models whose neutrino
sector involves only one triplet flavon plus one singlet flavon.
We have discussed two different minimal $A_{4}$ see-saw models, the ``usual'' one,
and a new ``alternative'' one. Working in the diagonal
charged lepton and right-handed neutrino mass basis,
we have shown that both these see-saw models satisfy the conditions of
FD, leading to two different neutrino mass sum rules which both predict closely spaced neutrino masses
with a normal or inverted neutrino mass ordering.
Despite the fact that they satisfy FD, we have seen that
the minimal $A_{4}$ models require some partial cancellations in order to obtain
an acceptable neutrino mass pattern. This is due to the fact that
the light physical neutrino mass eigenvalues $m_i$ are each non-trivial functions of the basic
parameters of the model, in particular the two flavon VEVs in the neutrino sector
of the minimal $A_{4}$ models. While this leads to some welcome
predictivity of the neutrino masses, it does mean that
partial cancellations between the two flavon VEVs are required to
achieve the desired hierarchy between the oscillation parameters
$\Delta m_{atm}^{2}$ and $\Delta m_{sol}^{2}$.

The most natural way to achieve FD without invoking
any cancellations would be to have three
triplet flavons $\tilde{\Phi}_1$, $\tilde{\Phi}_2$, $\tilde{\Phi}_3$, in the neutrino sector, whose
VEVs $\langle \tilde{\Phi}_1 \rangle$, $\langle \tilde{\Phi}_2 \rangle$, $\langle \tilde{\Phi}_3 \rangle$
are proportional to the three columns of the TBM matrix $\Phi_1$, $\Phi_2$, $\Phi_3$, respectively,
as in Eq.\ref{Phi0}. These three flavon VEVs
would then form the three columns of the Dirac neutrino mass matrix,
in the diagonal charged lepton and right-handed neutrino mass basis, as in Eq.\ref{FD2}. In such a type I see-saw model,
which we refer to as Natural Form Dominance (NFD), each light physical neutrino mass $m_i$ would
then be controlled by the VEV of a different flavon $\langle \tilde{\Phi}_i \rangle$,
allowing an arbitrary neutrino mass spectrum
to be achieved without requiring
any partial cancellations of parameters. In fact CSD \cite{King:2006np} just corresponds to
a special case of NFD corresponding to a strong neutrino mass hierarchy $|m_1|\ll |m_2| < |m_3|$,
in which the contribution of the flavon $\tilde{\Phi}_1$ is negligible (and hence may be
dropped or replaced by any other flavon) while $m_2$ is controlled by the flavon $\tilde{\Phi}_2$ and
$m_3$ is controlled by the flavon $\tilde{\Phi}_3$.
Examples of such CSD models, with $\tilde{\Phi}_1$ replaced by a flavon whose VEV alignment is approximately
proportional to $(0,0,1)^T$, have already been proposed
based on $SU(3)$, $\Delta_{27}$ \cite{deMedeirosVarzielas:2005ax},
$SO(3)$ \cite{King:2006np,King:2006me} and $A_4$ \cite{km}.
However NFD models with a general (not necessarily hierarchical) neutrino mass spectrum
involving all three flavons $\tilde{\Phi}_i$ have yet to be constructed.
This would represent an interesting
new direction in model building which goes beyond the minimal $A_4$ see-saw models considered here.

\section*{Acknowledgements}
We would like to thank the organizers of Neutrino 2008 and the Melbourne Theory Workshop 2008
where this work was started. M.-C.C. would also like to thank the hospitality of the Theory Group at the University of Southampton
where part of the work was completed.
The work of M.-C.C. was supported, in part, by the U.S.
National Science Foundation under Grant No. PHY-0709742.
SFK acknowledges partial
support from the following grants: STFC Rolling Grant
ST/G000557/1; EU Network MRTN-CT-2004-503369; EU ILIAS RII3-CT-2004-506222.

\end{document}